\documentclass[sigconf]{acmart}

\usepackage{listings}
\usepackage{algorithm}
\usepackage[noend]{algorithmic}

\usepackage{amsmath,amssymb}  
\usepackage{amsfonts}
\usepackage{graphicx}
\usepackage{enumitem}
\usepackage{amsthm}
\usepackage{xurl} 

\newtheorem{theorem}{Theorem}

\newtheorem{definition}{Definition}

\newcommand{\db}{\ensuremath{\mathcal{D}}}
\newcommand{\tb}{\ensuremath{d}}
\newcommand{\opset}{\ensuremath{\mathcal{O}}}
\newcommand{\op}{\ensuremath{o}}
\newcommand{\ops}{\ensuremath{S}}
\newcommand{\question}{\ensuremath{q}}
\newcommand{\col}{\ensuremath{c}}
\newcommand{\cols}{\ensuremath{C}}
\newcommand{\clu}{\ensuremath{g}}
\newcommand{\clus}{\ensuremath{G}}

\newcommand{\name}{TSO}

\AtBeginDocument{%
  }

\setcopyright{acmlicensed}
\copyrightyear{2018}
\acmYear{2018}
\acmDOI{XXXXXXX.XXXXXXX}

\acmConference[Conference acronym 'XX]{Make sure to enter the correct
  conference title from your rights confirmation emai}{June 03--05,
  2018}{Woodstock, NY}
\acmISBN{978-1-4503-XXXX-X/18/06}

\begin{document}

\title{Text to Query Plans for Question Answering on Large Tables}

\author{Yipeng Zhang}
\affiliation{%
  \institution{CSIRO Data61}
  \city{Clayton, VIC}
  \country{Australia}}
\email{yipeng.zhang@data61.csiro.au}

\author{Chen Wang}
\affiliation{%
  \institution{CSIRO Data61} 
  \city{Eveleigh, NSW}
  \country{Australia}}
\email{chen.wang@data61.csiro.au}

\author{Yuzhe Zhang}
\affiliation{%
  \institution{CSIRO Data61} 
  \city{Eveleigh, NSW}
  \country{Australia}}
\email{yuzhe.zhang@data61.csiro.au}

\author{Jacky Jiang}
\affiliation{%
  \institution{CSIRO Data61} 
  \city{Eveleigh, NSW}
  \country{Australia}}
\email{jack.jiang@data61.csiro.au}


\begin{abstract}
Efficient querying and analysis of large tabular datasets remain significant challenges, especially for users without expertise in programming languages like SQL. 
Text-to-SQL approaches have shown promising performance on benchmark data; however, they inherit SQL's drawbacks, including inefficiency with large datasets and limited support for complex data analyses beyond basic querying. We propose a novel framework that transforms natural language queries into query plans. Our solution is implemented outside traditional databases, allowing us to support classical SQL commands while avoiding SQL's inherent limitations. Additionally, we enable complex analytical functions, such as principal component analysis and anomaly detection, providing greater flexibility and extensibility than traditional SQL capabilities.
We leverage LLMs to iteratively interpret queries and construct operation sequences, addressing computational complexity by incrementally building solutions.
By executing operations directly on the data, we overcome context length limitations without requiring the entire dataset to be processed by the model. We validate our framework through experiments on both standard databases and large scientific tables, demonstrating its effectiveness in handling extensive datasets and performing sophisticated data analyses. 
\end{abstract}




\maketitle

\section{Introduction}

Tabular data is widely used for storing information. It plays a crucial role in data analytics in various fields such as finance, healthcare, scientific research, manufacturing and general business process management. With a two-dimensional representation format, tabular data makes it easy for users to manage structured information, enabling complex data analysis and insight extraction methods to be built on it. Complex data analytics require efficient data query and retrieval. Structured Query Language (SQL) is the standard for interacting with tables in relational databases. It allows users to perform operations like filtering, joining, and aggregating data with the support of underlying relational algebra. However, SQL has several disadvantages. First, it is not easily accessible to non-technical users, requiring knowledge of specific syntax and query structures. Second, SQL struggles to handle large datasets efficiently, especially when dealing with super-large tables that exceed database limitations. Complex partitioning and mapping are often required to support SQL executions on partitioned or distributed tables~\cite{xin2013shark, huai2014major, armbrust2015spark}. Third, SQL supports limited operations and cannot perform complex data analyses such as Principal Component Analysis (PCA), anomaly detection, or advanced pattern recognition.

Recent advances in large language models (LLMs) have enabled the approach of feeding the entire table into the LLM and generating answers to natural language queries~\cite{DBLP:conf/naacl/ZhangYL024, DBLP:journals/pacmmod/LiHYCGZF0C24}. However, this method encounters significant challenges due to the limited context length of LLMs.
To address these context limitations, one solution for table querying relies on compressing or truncating large tables to fit the context limits of the models~\cite{DBLP:journals/corr/abs-2307-03172}. These methods, however, often result in incomplete analyses and performance degradation, especially when working with complex datasets that contain thousands of columns and rows.
Moreover, even with efforts to compress input content or increase token limits, LLMs cannot effectively handle large tables, as they exceed the models' maximum input size~\cite{DBLP:conf/iclr/ChengX0LNHXROZS23, DBLP:conf/eacl/Chen23, DBLP:journals/corr/abs-2312-09039}.

There are efforts to use Text-to-SQL to address the problem, but many early work~\cite{androutsopoulos1995natural} were not widely used due to the difficulty of understanding natural language~\cite{li2014constructing}. As an alternative, researchers have explored LLM-based methods~\cite{DBLP:conf/nips/PourrezaR23, DBLP:conf/aaai/Li00023, DBLP:journals/corr/abs-2307-07306, DBLP:journals/pvldb/GaoWLSQDZ24} by providing only the table schema to the LLM. The model generates SQL queries that can be executed on the database according to the schema. This approach significantly alleviates the token limitation problem and has demonstrated good performance in recent studies. However, it still inherits SQL's inherent disadvantages: inefficiency with large datasets and inability to perform complex data analyses beyond basic querying. Here we particularly consider those datasets whose schemas do not fit into the context window of LLMs. 
While some efforts have attempted to teach LLMs to generate code directly or use APIs to overcome these limitations, problems related to scalability, efficiency, and the integration of complex analytical functions remain as a challenge for large-scale databases~\cite{li2024can}. 
While some efforts have attempted to teach LLMs to generate code directly or use APIs to overcome these limitations, problems related to scalability, efficiency, and the integration of complex analytical functions remain as a challenge for large-scale databases~\cite{li2024can}. 

To address the problems, we propose a novel approach that leverages the advantages of SQL while overcoming its limitations. Instead of converting natural language to SQL queries, \emph{we directly convert text-based queries to query plans corresponding to SQL-like queries of the text}. This provides flexibility to handle large data stored in the form of spreadsheets or CSV form, while remaining compatible with traditional relational databases. We create a set of SQL-like operators, such as selection by conditions, ordering, union, and joining, but implemented outside the constraints of traditional databases. These operators allow us to mimic SQL functionality without being hindered by database limitations or inefficiencies with large datasets, thus addressing the second disadvantage.
Another benefit of generating query plan directly is that operators are extensible and complex analytical functions required for specific tasks, such as dimension reduction, clustering, anomaly detection, and advanced pattern recognition can be easily integrated. This flexibility enables us to perform sophisticated data analyses directly within our framework, effectively solving the third disadvantage.

At the core of our approach is the problem of transforming a user's natural language query into a sequence of operations that retrieves and processes the relevant data from tabular datasets before returning the final answers to the user query. 
%
%
However, determining the optimal sequence of operations is a computationally challenging task. This problem is analogous to the classical NP-hard planning problem. The complexity arises from the vast number of possible operation sequences and the dependencies between operations. Therefore, it is impractical to exhaustively search for the optimal sequence in polynomial time.

To overcome this challenge, we employ an iterative approach that incrementally constructs the operation sequence based on the ReAct prompting framework~\cite{DBLP:conf/iclr/YaoZYDSN023}. By utilizing LLMs, we can understand the user's question and reason about the most appropriate next steps based on the current state of the data. LLMs are well-suited for this task due to their capabilities in natural language understanding and in-context learning~\cite{garg2022can, dong2022survey}. At each iteration, the model selects the next operation to apply, and the process continues until the final answer is obtained. In addition, we provide a multi-level table description generation mechanism to scale our method to handle large data tables with thousands of columns. 

We validate our approach through experiments on both traditional databases, Spider dataset~\cite{DBLP:conf/emnlp/YuZYYWLMLYRZR18}), and large scientific tables, agronomic dataset~\cite{newman2021multiple}. The experiments on the Spider dataset demonstrate that our solution performs well on traditional tabular data in Table QA tasks, even without utilizing any training data from the dataset, whereas the experiments on the agronomic dataset show that our solution is capable of handling super large tabular data under complex Table QA tasks.

\section{Related Work}

Supporting question-answering on tabular data has drawn increased attention as LLMs become increasingly powerful. Existing approaches can be broadly categorized into two categories: semantic parsing-based methods, and non-SQL-based question answering methodologies on structured data.


\subsection{Semantic Parsing-Based Methods}
Semantic parsing uses LLMs to transform natural language questions SQLs, and then run SQLs on data tables. Early solutions use encoder-decoder architectures to learn schema linking patterns~\cite{wang2020rat}. With the advent of LLMs, the accuracy of generating SQL has dramatically improved. These models have been continuously breaking records on benchmarks like Spider \cite{DBLP:conf/emnlp/YuZYYWLMLYRZR18}.

Studies \cite{DBLP:conf/aaai/LiHCQ0HHDSL23, DBLP:conf/aaai/Li00023, DBLP:conf/emnlp/QiTHW0ZWZL22, DBLP:conf/emnlp/ScholakSB21, DBLP:conf/slt/ZengPH22} focus on tuning or enhancing language models to improve performance in text-to-SQL tasks. Specifically, \citet{DBLP:conf/aaai/LiHCQ0HHDSL23} integrate graph-aware layers with a pre-trained T5 model to handle complex and multi-hop SQL queries, enhancing domain generalization. \citet{DBLP:conf/aaai/Li00023} introduce a ranking-enhanced encoding and skeleton-aware decoding framework within a seq2seq model, simplifying schema linking and enhancing SQL parsing. \citet{DBLP:conf/emnlp/QiTHW0ZWZL22} augment a Transformer seq2seq architecture with relation-aware self-attention, improving the model's capability to manage relational data effectively in text-to-SQL translations. \citet{DBLP:conf/emnlp/ScholakSB21} introduce incremental parsing techniques to constrain the decoding process of auto-regressive models, ensuring the generation of valid SQL by rejecting inadmissible tokens. \citet{DBLP:conf/slt/ZengPH22} propose a heuristic schema linking algorithm combined with a query plan model to rerank model-generated SQL queries.

Despite these advancements, semantic parsing-based methods inherit SQL's inherent limitations. SQL struggles with large tables due to database systems' constraints on the number of columns and inefficiencies when handling massive datasets. Additionally, SQL lacks support for complex data analyses beyond basic querying.

\subsection{Non-SQL approach on structured data}

There are many methods leveraging LLMs without relying on SQL for tabular data question answering. A naive approach is to feed the entire tabular data as a context directly into an LLM to answer user queries. 
The performance of this approach subjects to the context length limitation in LLMs. Many existing LLM-based solutions for table querying rely on compressing or truncating large tables to fit the context limits of models. These methods not only reduce the available data but also lead to incomplete analyses and performance degradation, especially when working with tables containing thousands of columns and rows~\cite{DBLP:journals/corr/abs-2307-03172}. Even worse, studies show that when dealing with large tabular data, the performance drops more than when handling normal textural content~\cite{DBLP:conf/iclr/ChengX0LNHXROZS23, DBLP:conf/eacl/Chen23, DBLP:journals/corr/abs-2312-09039}.

Other methods in this category include the use of code interpreters and the integration of LLMs with tools based on the ReAct framework~\cite{DBLP:conf/iclr/YaoZYDSN023}.


\subsubsection{Code Interpreter Methods}

Code interpreter methods enhance the coding abilities of LLMs to generate and execute code for data manipulation tasks. Notable works in this area include: 
\textit{SheetCopilot}~\cite{DBLP:conf/nips/LiSCLZ23} proposes an agent that interprets natural language tasks and controls spreadsheets using a set of atomic actions based on VBA, enabling LLMs to interact robustly with spreadsheet software.
Xue et al.~\cite{DBLP:conf/deem/LiD24} introduces a framework where users provide task instructions, and the system generates multiple candidate code snippets, ranks them, and selects the best one to solve the task, iteratively refining the code for unsolvable rows.
\textit{Binder}~\cite{DBLP:conf/iclr/ChengX0LNHXROZS23} presents a neural-symbolic framework that maps task inputs to programs combining SQL and LLM functionalities, using GPT-3 Codex for parsing and execution without task-specific training.

Direct code-generation is powerful and flexible for querying tabular data, but also suffers from uncertainty of generation without constraints. Troubleshooting is also difficult when results are unexpected.  

\subsubsection{ReAct with Tools Methods}

The ReAct~\cite{DBLP:conf/iclr/YaoZYDSN023} framework integrates LLMs with predefined tools to enhance accessibility and interpretability. Each operation can be easily understood by users, facilitating transparency in the data manipulation process. 
\textit{StructGPT}~\cite{DBLP:conf/emnlp/JiangZDYZW23} proposes an iterative reading-then-reasoning framework where LLMs utilize interfaces to interact with structured data, such as databases and knowledge graphs. For tables, it supports basic operations like extracting column names and sub-tables. While using tools and focusing on QA tabular data, it only processes single tables and offers a limited set of functions, restricting its ability to perform complex data manipulations or analyses across multiple tables.
\textit{TableLLM}~\cite{DBLP:journals/corr/abs-2403-19318} mainly focuses on enabling LLMs to manipulate tabular data embedded in documents and spreadsheets. It extracts tables and executes operations based on user instructions, primarily handling tasks like data filtering and chart generation.
\textit{ReAcTable}~\cite{DBLP:journals/pvldb/ZhangHFCDP24} is built upon the ReAct model and generates intermediate data representations to transform data into a more accessible format for answering questions. However, it relies on executing SQL and Python code, inheriting the disadvantages of both, including inefficiency with large datasets and potential security risks associated with code execution.

None of these methods focus on large tables that are commonly used in scientific research. 
\section{Methdology}

In this section, we begin by defining the problem of transforming a user's natural language query into a sequence of operations over tabular data. Due to the NP-hardness of finding an optimal solution, we propose a solution to solving complex Table QA tasks through a combination of large language models (LLMs) and a tree-structured planning framework, 
where raw tables serve as leaf nodes, intermediate results form internal nodes, and the final result is the root. We further introduce a three-level vector index system to facilitate efficient retrieval of columns among large tables. 

\subsection{Problem Definition}

Table QA tasks often require operations such as projection, sorting, grouping, aggregation, and joining across multiple tables. Determining the optimal sequence of operations to answer a user's query is computationally infeasible due to the NP-hardness of the problem, as we prove later.
To overcome the limitations of traditional methods, we formulate the problem as finding the correct sequence of operations that, when applied to the datasets, yields the desired result according to the user's query.

\begin{definition}[Problem Definition]\label{def:problem_definition}
Given a set of tabular data $\db = \{\tb_1, \tb_2, \cdots \tb_m \}$, a user query $\question$, and a set of operators on $\db$ denoted by $\opset = \{\op_1, \op_2, \cdots, \op_n\}$, our objective is to generate a logical plan $p \in P$ with $\opset_i \subseteq \opset$ so that $p(q, \db)$ yields a result that satisfies $\question$. Assuming the true answer to $\question$ is $y_q$, our objective is as below:

\begin{equation}
    \min_{p \in P} \quad L( p(q, \db), \, y_\question)
\end{equation}


\end{definition}

where each table contains columns $\tb_i.\cols$, each $\op_i$ is a fundamental operation (e.g., selection, projection, joining, aggregation, and advanced analytical functions); $P$ is all possible combinations of operators; A function $f(\db, \question, \opset)$ produces $p$; $\ops(\db)$ denotes the application of the sequence $\ops$ to the datasets $\db$, resulting in the processed data; $L$ is a loss function that quantifies how well the result satisfies the user's query. For instance, it could be defined based on the logical correctness of $p$ to answer \question \ on $\db$ or the distance between the answer produced by $p$ and the true answer.

\subsection{Hardness of the Problem}

\begin{theorem}\label{theorem:nphard}
    Finding the optimal sequence of operations is an NP-hard problem.
\end{theorem}

Determining the optimal sequence $p$ that minimize $L( p(\question, \db), \, y_\question)$ is computationally challenging. This problem is analogous to the classical planning problem in artificial intelligence, which is known to be NP-hard due to the combinatorial explosion of possible operation sequences and dependencies between operations. The proof is shown in Appendix~\ref{sec:app_nphard}. Therefore, it is impractical to exhaustively search for the optimal sequence in polynomial time.

\subsection{Tree Structure Representation}

\begin{figure}
    \centering
    \includegraphics[width=1\linewidth]{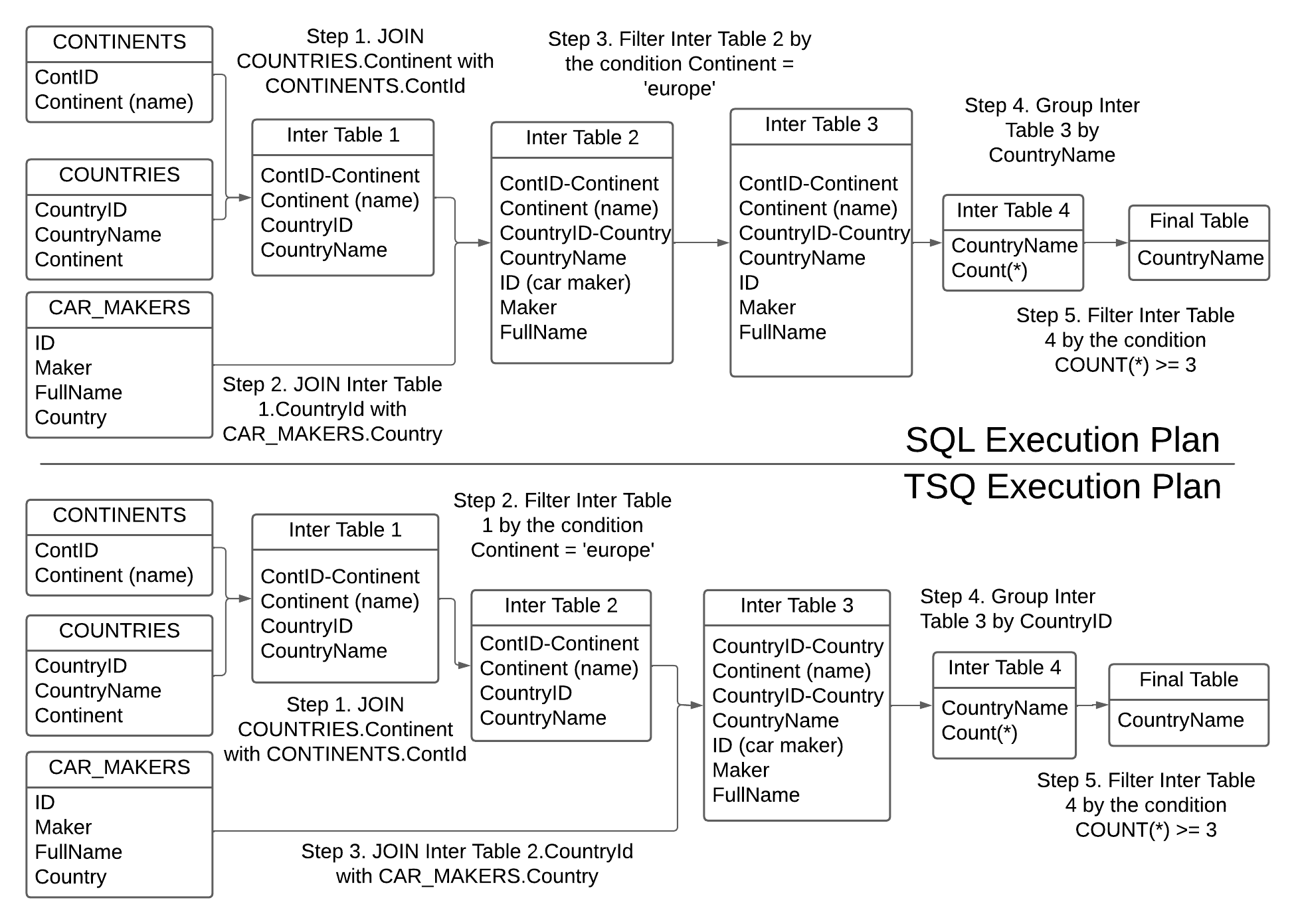}
    \caption{The Example of Tree Structure Planning}
    \label{fig:tree_structure}
\end{figure}

To effectively address the complexity of Table QA tasks, we propose a tree-structured plan that represents the sequence of operations as a computational graph. This structure allows us to decompose the problem into manageable steps and leverage the capabilities of LLM in an iterative manner. More importantly, research~\cite{DBLP:books/aw/AhoSU86, DBLP:conf/ijcai/NauCLM99} has demonstrated that the tree-structured plan and the sequence of operations can be transformed into one another without loss of information or functionality. This bidirectional transformation is the foundation of our solution, as it ensures that a feasible tree-structured plan is the solution of our problem.


Our tree-structured plan consists of the following components:

\begin{itemize} [leftmargin=*]
\item \textbf{Leaves (Initial Tables)}: The independent tabular datasets $\tb_i \in \db$ are represented as the leaf nodes of the tree. Each leaf node corresponds to a raw dataset serving as input to the process. 

\item \textbf{Intermediate Nodes (intermediate results)}: Each intermediate node represents an intermediate result, which executes an operation $\op_i \in \opset$ on one or more child nodes (initial tables or intermediate results) to produce a new parent node.

\item \textbf{Root Node (Final Result)}: The root node represents the final answer to the user's query, obtained after performing all necessary operations on the data. 
\end{itemize}

Figure~\ref{fig:tree_structure} illustrates how our system leverages tree-structure planning to progressively generate the final result through a sequence of operations on tabular data. Both plans aim to answer the query: "Which countries in Europe have at least 3 car manufacturers?" This question is from the Spider Dataset~\cite{DBLP:conf/emnlp/YuZYYWLMLYRZR18}, and the SQL is the ground truth of this query. This question involves three tables, CONTINENTS, COUNTRIES, and CAR\_MAKERS, which are also the leaves in our Tree structure solution.

The SQL-based plan begins by joining the COUNTRIES and CONTINENTS tables using the `Continent' and `ContId' fields to combine country and continent information. It then joins this result with the CAR\_MAKERS table on the `Country' field, adding car manufacturer details such as `ID', `Maker', and `FullName'. After filtering to include only countries in Europe, the data is grouped by `CountryName', counting the number of car manufacturers per country. The system filters the groups to keep only those with at least 3 manufacturers and finally selects the `CountryName' column to produce the list of countries in Europe with three or more car manufacturers.



Our plan executs in a more efficient way by filtering countries by continent earlier in the execution. After joining the COUNTRIES and CONTINENTS tables to create Inter Table 1 with combined country and continent data, we immediately filter this intermediate table to retain only European countries (Continent = 'Europe'). This early filtering reduces the dataset before the join with CAR\_MAKERS, resulting in a smaller Inter Table 3 that includes only relevant data for European countries.
Following this join, the remaining steps in our plan are similar to those in the SQL execution plan: we group Inter Table 3 by CountryID, count the number of car manufacturers per country to get Inter Table 4. Finally, we filter to keep only those with at least three manufacturers, and extract CountryName into Final Table. This optimized approach minimizes intermediate data size and avoids unnecessary operations, making the execution more efficient.



This process demonstrates the iterative nature of tree-structure planning, where operations are sequentially applied, and the LLM evaluates intermediate results to decide the next step, discarding irrelevant data when necessary to refine the path toward the final result. 
Overall, the tree-structured plan offers four advantages:

\paragraph{Learning Relational Tasks}

Constructing the tree involves discovering and applying appropriate relational operations that integrate various data sources to derive the final result. It effectively captures the relational reasoning required in complex Table QA tasks.

\paragraph{Sequential Generation of the Computational Graph}

The tree structure can be linearized, generating the computational graph sequentially from the leaves to the root. This linearization enables us to process operations step by step, applying one operation at a time and progressively building towards the final result.

\paragraph{Feasibility with Large Language Models}

By linearizing the process, we make it manageable for LLMs to handle. The LLM can focus on selecting and applying one operation at a time rather than generating the entire operation sequence in one step, which would be impractical due to computational constraints. (might combine with the previous one)

\paragraph{Ability to Backtrace}

By maintaining all intermediate nodes (tables), we enable the framework to backtrack to previous operations if the LLM realizes that a certain path does not lead towards the desired outcome. This enhances the robustness of the solution by allowing corrections and adjustments, as any node above the leaves represents the result of a sequence of operations applied thus far.

\subsection{The Architecture}

\begin{figure}
    \centering
    \includegraphics[width=1\linewidth]{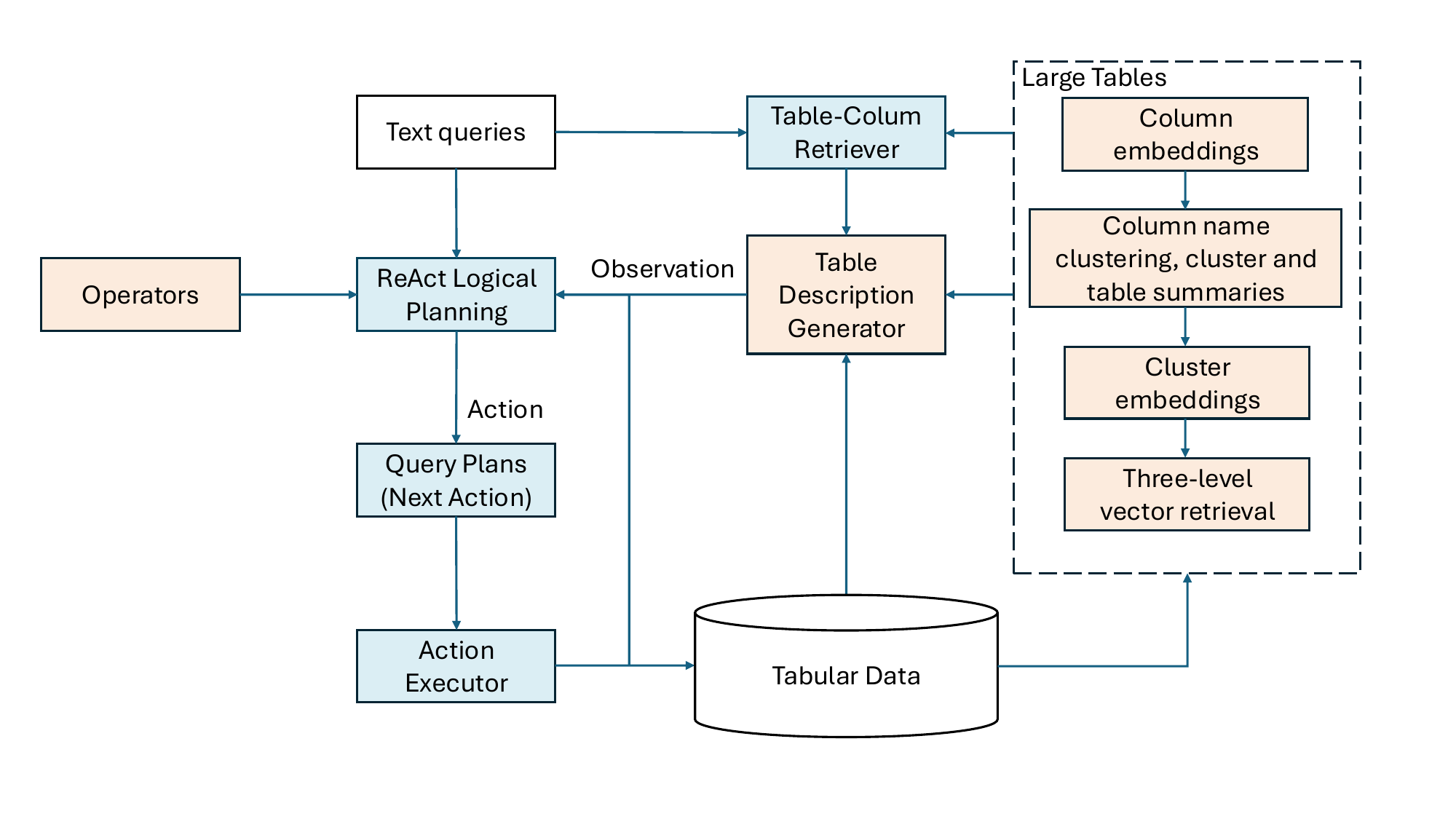}
    \caption{The Architecture of Tree-Driven Sequential Operation QA System (TSO)}
    \label{fig:system_structure}
\end{figure}

Now, we are ready to introduce our solution, a \textbf{T}ree-Driven \textbf{S}equential \textbf{O}peration QA System (\name) that seamlessly combines observing the tree-structure state and reasoning the next operation. The ReAct (Reasoning and Acting) framework provides such an integration, allowing LLMs to not only interpret and reason user queries but also to execute actions through predefined tools. By adopting the ReAct framework, we can decompose complex queries into manageable tasks, execute them efficiently, and iteratively refine the results to satisfy the user's query. Figure~\ref{fig:system_structure} shows the overall system structure.
At the center of our system is a supervisor agent, an LLM responsible for interpreting user queries and orchestrating the execution of tasks using available tools. The supervisor agent operates through an iterative loop comprising the following steps:

\begin{enumerate} \item \textbf{Thought}: 
The supervisor agent assesses the current state of the tree, including all existing intermediate nodes and their results. Using the tree structure, the agent determines which relational operations are required next to move closer to the final result. It identifies dependencies and potential operations based on the current intermediate results.

\item \textbf{Action}: 
The agent chooses the most appropriate operation $\op_i \in \opset$ to apply to one or more child nodes. When the agent applies an operation, it adds a new intermediate node to the tree. This operation combines data from the relevant child nodes (which could be initial tables or previous intermediate results) to create a new parent node.

\item \textbf{Observation}: After performing the operation, the agent examines the result, which is the new intermediate node, and evaluates whether it helps answer the user's query. The LLM observes the current state by reviewing all tables and their descriptions generated by the table description generator.

\item \textbf{Backtracking}: 
If the observation indicates that the current path is not leading towards the desired outcome, the agent can backtrack by revisiting previous nodes and reconsidering alternative operations. The ability to backtrace leverages the tree's hierarchical structure, allowing the agent to navigate to different branches and explore alternative sequences of operations. This step is optional and is integrated with the Action step.
\end{enumerate}

As the iterative loop progresses, each operation adds a new node to the tree, building the computational graph step-by-step from the leaves to the root.

\noindent\textbf{Toolset Description. }
We have developed a comprehensive set of tools within the ReAct framework to perform various data manipulation and analysis tasks. These tools are designed to handle large datasets and can be easily extended to include new functionalities. The key tools in our system include:

\begin{itemize}[leftmargin=*]
    \item \textit{Data Loading Tool}: Reads tabular files (e.g., CSV, Excel) into a DataFrame for processing. 
    \item \textit{Data Sampling Tool}: Retrieves sample rows from a DataFrame to provide an overview of the data. 
    \item \textit{Aggregation Tool}: Performs aggregation operations such as sum, mean, count, and max on selected columns. 
    \item \textit{Grouping Tool}: Groups data by specified columns and optionally orders the results based on other columns. 
    \item \textit{Dataframe Introduction Tool}: Generates summaries of DataFrames, including column indices and descriptions. 
    \item \textit{Join Tool}: Merges two DataFrames based on specified join types and columns. 
    \item \textit{Sorting Tool}: Sorts data within a DataFrame according to specified columns and order. 
    \item \textit{Selection and Filtering Tool}: Selects specific columns and filters rows based on given conditions. 
    \item \textit{Set Operations Tool}: Executes set operations (e.g., union, intersection) between two DataFrames. 
    \item \textit{Advanced Analysis Tools}: Performs complex analyses such as Principal Component Analysis (PCA), anomaly detection, PythonREPLTool, and value prediction. 
\end{itemize}

These tools are implemented in Python and can be easily extended or modified to accommodate additional functions, enhancing the system's adaptability to various analytical needs.

\subsection{Large Tables Understanding}

In large-scale scientific data analysis, researchers often deal with datasets that have thousands of columns spread across multiple tables. This massive scale creates significant challenges when the ReAct model is trying to understand all tables. Feeding the entire dataset or schema is impractical due to token limits. To handle this, we create multi-level vector indexes for columns, clusters, and tables. By converting descriptions of these elements into vector formats (Algorithm~\ref{alg:build_vector_stores}), we can quickly compare the user’s query with the data components (Algorithm~\ref{alg:query_relevant_columns}). This helps us find and provide only the necessary columns to the LLM, bypassing the context length issue while ensuring the system remains scalable and efficient.

\begin{algorithm}[t]
\caption{Build Vector Stores}
\small
\label{alg:build_vector_stores}
\begin{algorithmic}[1]
\STATE{\bf Initialization:} $\db = \{\tb_1, \tb_2, \dots, \tb_m\}$, $\mathbf{l}_\cols$, $\mathbf{l}_\clus$, $\mathbf{l}_\db$
\FORALL{tables $\tb_i \in \db$}
    \FORALL{columns $\col_k \in \tb_i.\cols$}
        \STATE $\col_k.{\text{Des}} \leftarrow \text{LLM\_Describe}(\col_k)$
        \STATE $\mathbf{l}_{\col_k} \leftarrow \text{Embed}(\col_k.{\text{Des}})$
        \STATE Store $(\text{col\_id}: \col_k.\text{id}, \mathbf{l}_{\col_k}, \col_k.\text{Des})$ in $\mathbf{l}_{\cols}$
    \ENDFOR
    \STATE Cluster $\tb_i.\cols$ to obtain clusters $\{\clus_i\}$ based on $\mathbf{l}_{\col_k}$
    \FORALL{clusters $\clu_j$ in $\clus_i$}
        \STATE $\clu_j.{\text{Des}} \leftarrow \text{LLM\_Describe}(\{\col_k.{\text{Des}} \mid \col_k \in \clu_j\})$
        \STATE $\mathbf{l}_{\clu_j} \leftarrow \text{Embed}(\clu_j.{\text{Des}})$
        \STATE Store $(\text{cluster\_id}: \clu_j.\text{id}, \mathbf{l}_{\clu_j}, \clu_j.{\text{Des}})$ in $\mathbf{l}_{\clus}$
    \ENDFOR
    \STATE $\tb_i.{\text{Des}} \leftarrow \text{LLM\_Describe}(\{\clu_j.{\text{Des}} \mid \clu_j \in \clus_i\})$
    \STATE $\mathbf{l}_{\tb_i} \leftarrow \text{Embed}(\tb_i.{\text{Des}})$
    \STATE Store $(\text{table\_id}: \tb_i.\text{id}, \mathbf{l}_{\tb_i}, \tb_i.\text{Des})$ in $\mathbf{l}_{\db}$
\ENDFOR
\STATE{\bf Return:} $\mathbf{l}_\cols$, $\mathbf{l}_\clus$, $\mathbf{l}_\db$

\end{algorithmic}
\end{algorithm}

Algorithm~\ref{alg:build_vector_stores} constructs the vector stores necessary for efficient query processing by embedding the semantic descriptions of columns, clusters, and tables in a hierarchical structure. The algorithm begins by iterating through each table $\tb$ in the database $\db$. For each table, it processes its columns by iterating over each column $\col$ (Lines 3–4). A human-readable description of each column is generated using an LLM, which captures the semantic meaning of the column data. The column description is then embedded into a vector $\mathbf{l}{\col}$ (Line 5), enabling numerical similarity computations. The column embeddings and descriptions are stored in the vector store $\mathbf{l}\cols$ using column IDs as keys (Line 6).

Once the columns have been processed, the algorithm clusters the columns based on their embeddings (Line 7). For each resulting cluster $\clu_j$, the LLM summarizes the descriptions of the columns within the cluster to generate a cluster description (Lines 8–9). This cluster description is then embedded into a vector $\mathbf{l}{\clu_j}$ (Line 10), and both the cluster embeddings and descriptions are stored in the vector store $\mathbf{l}\clus$ using cluster IDs as keys (Line 11).

After processing the clusters, the algorithm generates a description for the entire table by summarizing the descriptions of the clusters within it (Lines 12–13). This table description is embedded into a vector $\mathbf{l}{\tb}$, and the table embeddings and descriptions are stored in the vector store $\mathbf{l}\db$ using table IDs as keys (Line 14).

Finally, in Line 15, the algorithm returns the vector stores $\mathbf{l}\cols$, $\mathbf{l}\clus$, and $\mathbf{l}_\db$, each containing embeddings and descriptions for columns, clusters, and tables. This hierarchical embedding structure allows for efficient query processing by enabling similarity searches at multiple levels (columns, clusters, and tables), ensuring scalability when dealing with large datasets.

\begin{algorithm}[t]
\caption{Query Relevant Columns}
\small
\label{alg:query_relevant_columns}
\begin{algorithmic}[1]
\STATE{\bf Initialization:} $\question$, $\mathbf{l}_\cols$, $\mathbf{l}_\clus$, $\mathbf{l}_\db$, $\theta_t$, $\theta_c$, $\theta_l$

\STATE $\mathbf{q} \leftarrow \text{Embed}(\question)$

\FORALL{tables $\tb$ with embeddings $\mathbf{l}_{\tb}$}
    \STATE $s_{\tb} \leftarrow \text{sim}(\mathbf{q}, \mathbf{l}_{\tb})$
\ENDFOR
\STATE $\mathcal{T} \leftarrow \{ \tb \mid s_{\tb} \geq \theta_t \}$

\FORALL{tables $\tb \in \mathcal{T}$}
    \IF{Validate$(\question, \text{LLM\_Describe}(\tb))$ is relevant}
        \FORALL{clusters $\clu_j$ in $\tb$ with embeddings $\mathbf{l}_{\clu_j}$}
            \STATE $s_j \leftarrow \text{sim}(\mathbf{q}, \mathbf{l}_{\clu_j})$
        \ENDFOR
        \STATE $\mathcal{C} \leftarrow \{ \clu_j \mid s_j \geq \theta_c \}$

        \FORALL{clusters $\clu_j \in \mathcal{C}$}
            \IF{Validate$(\question, \text{LLM\_Describe}(\clu_j))$ is relevant}
                \STATE $\cols^* \leftarrow \text{Validate}(\question, \text{LLM\_Describe}(\clu_j))$
                \IF{$\cols^* = \text{yes}$}
                    \STATE $\cols_{rel} \leftarrow \cols_{rel} \cup \clu_j.\cols$
                \ELSE
                    \FORALL{columns $\col_k \in \clu_j.\cols$ with embeddings $\mathbf{l}_{\col_k}$}
                        \STATE $s_k \leftarrow \text{sim}(\mathbf{q}, \mathbf{l}_{\col_k})$
                    \ENDFOR
                    \STATE $\cols_{cand} \leftarrow \{ \col_k \mid s_k \geq \theta_l \}$
                    \FORALL{columns $\col_k \in \cols_{cand}$}
                        \IF{Validate$(\question, \text{LLM\_Describe}(\col_k))$ is relevant}
                            \STATE $\cols_{rel} \leftarrow \cols_{rel} \cup \{ \col_k \}$
                        \ENDIF
                    \ENDFOR
                \ENDIF
            \ENDIF
        \ENDFOR
    \ENDIF
\ENDFOR
\RETURN $\cols_{rel}$
\end{algorithmic}
\end{algorithm}

Algorithm~\ref{alg:query_relevant_columns} retrieves the columns relevant to a user’s question by leveraging the vector embeddings of tables, clusters, and columns, stored in hierarchical vector stores. The algorithm begins by embedding the user’s natural language query $\question$ into a vector $\mathbf{q}$ using the embedding function (Line 2). The query embedding is then compared with the vector embeddings of all tables $\mathbf{l}{\tb}$ in the database to compute similarity scores $s{\tb}$ (Lines 3–4). Tables with similarity scores above a threshold $\theta_t$ are selected as candidate tables (Line 5). These tables are stored in $\mathcal{T}$ for further evaluation.

For each candidate table $\tb$ in $\mathcal{T}$, the algorithm first validates whether the table is relevant to the user’s question by checking the semantic description generated by the LLM (Line 7). If the table is relevant, the algorithm proceeds by calculating similarity scores between the query embedding and the cluster embeddings $\mathbf{l}_{\clu_j}$ within the table. Clusters with similarity scores above a threshold $\theta_c$ are selected and stored in $\mathcal{C}$ (Lines 8-10).

From Lines 12-15, the algorithm validates each selected cluster to determine if its semantic description aligns with the user’s question. If the entire cluster is relevant, the algorithm includes all columns from the cluster in the set of relevant columns $\cols_{rel}$, otherwise, the algorithm evaluates the relevance of individual columns within the cluster by computing similarity scores between the query embedding and column embeddings $\mathbf{l}{\col_k}$ (Lines 16–18). Columns with similarity scores above a threshold $\theta_l$ are considered as candidates and validated using their semantic descriptions (Lines 19–22). If validated, these columns are added to $\cols{rel}$.

After processing all relevant clusters in all relevant tables, the algorithm returns the set of relevant columns $\cols_{rel}$ (Line 23). This hierarchical process ensures that the most relevant columns are selected based on their similarity to the user’s query and their alignment with the query’s semantic intent.

\section{Experiment}

In this section, we provide experimental results on two practical datasets, the Spider dataset~\cite{DBLP:conf/emnlp/YuZYYWLMLYRZR18} and the agronomic dataset~\cite{newman2021multiple}. The experiments on the Spider dataset demonstrate that our solution performs well on traditional tabular data in Table QA tasks, even without utilizing any training data from the dataset. Moreover, we will discuss the deficiency of the Spider Data. The experiments on the agronomic dataset show that our solution is capable of handling super large tabular data under complex Table QA tasks.

\subsection{Dataset}

Spider~\cite{DBLP:conf/emnlp/YuZYYWLMLYRZR18} is a widely recognized benchmark dataset for Text-to-SQL tasks. It contains 8,659 instances in the training split and 1,034 instances in the development split across 200 databases. Each instance consists of a natural language question about a specific database and its corresponding SQL query. In this paper, we use the development split (Spider-dev) only for evaluation purposes, as we do not utilize the training data for model training. This approach allows us to assess the generalization ability of our solution without any fine-tuning of the training data.

The agronomic dataset~\cite{newman2021multiple} is a comprehensive dataset that contains results of the crop yield experiments from Australia. The dataset serves for a purpose of understanding crop growth and development under varying environmental and meteorological conditions. The dataset collects trial results from 2008 to 2018. The dataset is structured around various data domains (DOMs), each providing unique insights into different aspects of crop trials. 
The dataset captures time-series data, with many features recorded at multiple time intervals relative to the planting date. In total, the agronomic dataset comprises 266,033 records and 8,058 features, making it a substantial resource for evaluating the scalability and effectiveness of our solution on large-scale, complex datasets. The dataset represents a typical trend of tabular data organisation in the ``big data'' era in scientific domains. For more details, please see Appendix~\ref{sec:app_GRDC}

\subsection{Experiment Setting}

\noindent \textbf{Metric.} For the Spider dataset, since our solution outputs DataFrames instead of SQL queries, we execute the ground truth SQL queries and compare their results with our DataFrames. We use the Exact Match (EM) accuracy metric, which measures the proportion of queries where our result exactly matches the ground truth. This metric directly assesses the correctness of the data retrieved, regardless of how the query is written. We report EM accuracy across different query hardness levels (Easy, Medium, Hard, and Extra Hard) as defined in the Spider dataset.

For the agronomic dataset, because it doesn't have predefined questions, we designed 20 queries: 10 easy, 5 medium, and 5 hard. The easy queries involve straightforward operations like selection and ordering; medium queries require multi-step operations involving multiple columns; hard queries involve complex operations like principal component analysis (PCA) or anomaly detection. We manually obtained the ground truth answers using standard data processing tools and compared them with our solution's outputs. We report EM accuracy for each difficulty level to assess our solution's performance across varying query complexities.

\noindent \textbf{Baselines.} For the Spider dataset, we compare our method \name with DIN-SQL~\cite{DBLP:conf/nips/PourrezaR23}, as only this work provides the per-hardness EM accuracy metrics, and it ranks third on the Spider Execution with Values leaderboard, only 1.501\% behind the top method. DIN-SQL provides per-hardness EM accuracy metrics, making it a strong benchmark for our evaluation. For the agronomic dataset, since there are no existing baselines for our specific task, we focus on evaluating our solution's performance on the designed queries and discuss its effectiveness in handling complex, real-world data.







\subsection{Scientific Tabular Data}

We conducted experiments on the agronomic dataset using 20 carefully designed questions of varying difficulty levels—10 easy, 5 medium, and 5 hard—to evaluate the effectiveness of our solution. For the prediction task, we consider a prediction "correct" if its percentage error is within 10\%. The percentage error is calculated as $\frac{1}{y_\question}|p(q, \db) - y_\question|*100\%$. Our method correctly answers 16 out of 20 queries. Due to space limitations, we only show four of the questions where our solution did not provide the correct answer to show the limitations of our approach. Noting that, the column size exceeds many existing work's processing capability. For the full list, please refer to Appendix~\ref{sec:app_GRCD_question} and Table~\ref{table:GRDC}. 

\vspace{.3em}
\noindent \textbf{Question 7: List the different crop rotations recorded one year before planting.} This question is to retrieve the column METADom\_Crop\_rotation\_minus\_1\_sub\_cropWheat, which indicates the presence of wheat in the crop rotation one year before planting. However, the generated description for this column was misleading to the Table-Column Retriever. The description stated: "This column indicates the presence of wheat within the crop rotation system one day before planting (day -1)." The misinterpretation is because the description suggested a time frame of one day before planting, whereas the column represents one year prior. This discrepancy led the Table-Column Retriever to incorrectly assess the relevance of the column to the user's query, resulting in an incorrect answer.

\vspace{.3em}
\noindent \textbf{Question 15: How does cumulative evapotranspiration over the first 80 days after planting relate to grain yield?} This question is to relative to column PHENDom\_X1000.grain.weight\_2, 3, and 6. However, the question was ambiguous, leading the language model to return only the first relevant column, PHENDom\_X1000.grain.weight\_2, instead of all three. This indicates a limitation in handling ambiguous queries and suggests that providing more explicit instructions or incorporating clarification steps could improve the results.

\vspace{.3em}
\noindent \textbf{Questions 19: For trials with high waterlogging, assess whether soil properties contribute to the condition; Question 20: Use machine learning to predict the breeder based on phenotypic and environmental data.} The two questions are complex analytical tasks that likely involve advanced computations or multi-step data processing. During the search for the answers, the supervisor agent decided to utilize the \textit{PythonREPLTool} to execute Python code to solve the problems. However, the agent became trapped in debugging code errors and was unable to find a solution within the maximum iteration limit.

Despite these challenges, the overall performance indicates that, with the proper tools, our solution effectively interprets and processes user queries over large tabular datasets. Future work may focus on improving ambiguity resolution, enhancing description accuracy, and developing more robust code generation techniques to further increase the system's capabilities.

\subsection{Spider Dataset}


We evaluated our solution on the Spider dataset to assess its performance in interpreting natural language queries over complex databases. The results are shown in Table~\ref{table:spider}. Our solution has two versions: TSO and TSO$^{-}$. The difference is that TSO has access to the database schema, while TSO$^{-}$ does not. We have the following observations:

Our solution TSO achieved the second-best overall score, performing well across all difficulty levels. Notably, TSO got the best results in the \emph{Easy} and \emph{Extra Hard} categories, showing its effectiveness in handling both simple and complex queries. In contrast, TSO$^{-}$ performed worse than TSO. This shows the importance of providing the schema to the solution. Knowing the schema helps the system accurately map user queries to the relevant tables and columns, especially in complex databases with many tables.

The result shows that GPT models perform better in handling complex Table QA tasks. This trend matches the Spider leaderboard, where GPT-4-based solutions are among the top performers. The advanced reasoning and language understanding of GPT models help our solution perform well across various query difficulties.

\begin{table}[]
\caption{Experimental Results on the Spider Dataset. The best results are highlighted in bold, and the second-best results are underlined.}
\label{table:spider}
\small
\begin{tabular}{lllllll}
\hline
Baseline          & Model              & All            & Easy           & Medium         & Hard           & Extra          \\ \hline
TSO$^-$               & GPT-4o-mini        & 33.85          & 53.15          & 35.71          & 22.37          & 11.11          \\
TSO$^-$               & GPT-4o             & 45.75          & 64.36          & 47.87          & 44.59          & 32.53          \\
TSO$^-$               & Llama3.1 & 17.81          & 50.00          & 14.89          & 2.70           & 0.00           \\
TSO               & GPT-4o-mini        & 48.49          & 64.52          & 55.16          & 25.19          & 19.70          \\
TSO               & GPT-4o             & \underline {70.34}    & \textbf{93.02} & 69.92          & \underline {62.26}    & \textbf{47.69} \\
TSO               & Llama3.1 & 38.10          & 46.67          & 43.55          & 25.93          & 31.82          \\
DIN-SQL1 & GPT-4              & \textbf{74.20} & \underline {91.10}    & \textbf{79.80} & \textbf{64.90} & \underline {43.40}    \\
DIN-SQL2  & GPT-4              & 67.40          & 86.70          & \underline {73.10}    & 59.20          & 31.90          \\ \hline
\end{tabular}
\end{table}

\noindent\textbf{Discussion on Spider Dataset. } We found that some discrepancies were not due to errors in our solution but were caused by issues within the ground truth itself. From the six distinct types of errors including inconsistencies in query formulation, improper handling of NULL values, unjustified semantic inferences, misinterpretations of domain-specific terminology, data type mismatches, and ambiguities in query interpretation, we illustrate how these issues can significantly impact the accuracy and reliability of QA systems. These errors not only affected the evaluation of our solution's performance but also underscored how such issues can substantially influence the outcomes of QA models. 

\paragraph{1. Inconsistent Use of the DISTINCT Clause}

\noindent Question 1: "Find the first name and age of students who have a pet."
\begin{lstlisting}[frame=single, breaklines, keepspaces, basicstyle=\footnotesize, xleftmargin=0pt]
SELECT DISTINCT T1.fname, T1.age FROM student AS T1 
JOIN has_pet AS T2 ON T1.stuid = T2.stuid;
\end{lstlisting}

\noindent Question 2: "List all singer names in concerts in year 2014."
\begin{lstlisting}[frame=single, breaklines, keepspaces, basicstyle=\footnotesize]
SELECT T2.name FROM singer_in_concert AS T1
JOIN singer AS T2 ON T1.singer_id = T2.singer_id
JOIN concert AS T3 ON T1.concert_id = T3.concert_id
WHERE T3.year = 2014;
\end{lstlisting}

\noindent \textbf{Discussion. }
The ground truth SQL queries display an inconsistency in handling duplicate results due to the selective use of the DISTINCT clause. While the first query removes duplicates to present unique combinations of student names and ages, the second query does not eliminate duplicates of singer names. This inconsistency can lead to unreliable evaluations of QA models, as the presence or absence of duplicates affects the correctness of the output. For consistent and accurate results, similar queries should uniformly apply the DISTINCT clause when duplicates are possible.

\paragraph{2. Improper Treatment of NULL Values in Calculations}

\noindent Question: "For each zip code, what is the average mean temperature for all dates that start with '8'?"

\begin{lstlisting}[frame=single, breaklines, keepspaces, basicstyle=\footnotesize]
SELECT zip_code  ,  avg(mean_temperature_f) FROM weather 
WHERE date LIKE "8/%" GROUP BY zip_code
\end{lstlisting}

\noindent \textbf{Discussion. }
The ground truth SQL may incorrectly compute the average temperature by misinterpreting NULL values as zeros. In data science and SQL standards, NULL values should be excluded from aggregate functions like AVG(). Treating NULL as zero introduces bias and leads to inaccurate results. The error highlights the need for careful handling of missing data to ensure the validity of statistical computations in SQL queries.

\paragraph{3. Assumptive Semantic Substitution}

\noindent Question: "Show the medicine names and trade names that cannot interact with the enzyme with product 'Heme'."
\begin{lstlisting}[frame=single, breaklines, keepspaces, basicstyle=\footnotesize]
SELECT name, trade_name FROM medicine EXCEPT
SELECT T1.name, T1.trade_name FROM medicine AS T1
JOIN medicine_interaction AS T2 ON T2.medicine_id = T1.id
JOIN enzyme AS T3 ON T3.id = T2.enzyme_id
WHERE T3.product = 'Protoporphyrinogen IX';
\end{lstlisting}

\noindent \textbf{Discussion. }
The ground truth query makes an unwarranted inference by replacing the user-provided term 'Heme' with 'Protoporphyrinogen IX'. This substitution is not justified within the given context and disregards the user's explicit input. Such semantic assumptions can lead to incorrect query results and misinterpretation of the user's intent. Accurate SQL generation should adhere strictly to the user's specified terms unless additional context or clarification is provided.

\paragraph{4. Terminology Misalignment in Domain Concepts}

\noindent Question: "Find the model of the car whose weight is below the average weight."
\begin{lstlisting}[frame=single, breaklines, keepspaces, basicstyle=\footnotesize]
SELECT T1.model FROM CAR_NAMES AS T1
JOIN CARS_DATA AS T2 ON T1.MakeId = T2.Id
WHERE T2.Weight < (SELECT AVG(Weight) FROM CARS_DATA);
\end{lstlisting}

\noindent{Ground-Truth:}
\begin{lstlisting}[frame=single, breaklines, keepspaces, basicstyle=\footnotesize]
Model
toyota
plymouth
...
\end{lstlisting}

\noindent \textbf{Discussion. }
The ground truth SQL misinterprets the term "model" by returning car makes instead of models. In the automotive domain, the make is the manufacturer (e.g., Toyota), and the model is the specific vehicle line (e.g., Camry). This misalignment leads to incorrect results that do not satisfy the user's query. Precise understanding of domain-specific terminology is crucial for generating accurate SQL queries that reflect the user's intent.

\paragraph{5. Incorrect Data Type Handling in Numeric Comparisons}

\noindent Question: "What is the number of the cars with horsepower more than 150?"
\begin{lstlisting}[frame=single, breaklines, keepspaces, basicstyle=\footnotesize]
SELECT COUNT(*) FROM CARS_DATA
WHERE horsepower > 150;
\end{lstlisting}

\noindent \textbf{Discussion. }
The ground truth SQL fails to account for the data type of the horsepower column, which is stored as a string. Performing numerical comparisons on string data can yield erroneous results due to lexicographical ordering (e.g., '200' is considered less than '80' because '2' comes before '8'). To ensure accurate comparisons, the query should cast the horsepower column to a numeric data type before applying the comparison operator. This oversight highlights the importance of data type considerations in SQL queries involving numerical operations.

\paragraph{6. Ambiguous Reference to Entity Identifiers}

\noindent Question: "What are all the makers and models?"
\begin{lstlisting}[frame=single, breaklines, keepspaces, basicstyle=\footnotesize]
SELECT Maker, Model FROM MODEL_LIST;
\end{lstlisting}

\noindent
\begin{minipage}{0.22\textwidth}
\noindent{Ground-Truth:}
\begin{lstlisting}[frame=single, breaklines, keepspaces, basicstyle=\footnotesize]
Maker   Model
1       amc
2       audi
3       bmw
...     ...
\end{lstlisting}
\end{minipage}
\hfill
\begin{minipage}{0.22\textwidth}
\noindent{Our Result:}
\begin{lstlisting}[frame=single, breaklines, keepspaces, basicstyle=\footnotesize]
Maker        Model
amc          amc
volkswagen   audi
volkswagen   volkswagen
...          ...
\end{lstlisting}
\end{minipage}

\noindent \textbf{Discussion. }
The ground truth SQL returns maker IDs instead of maker names, which may not align with the user's expectation of obtaining human-readable information. In cases where identifiers can represent multiple entities (e.g., IDs vs. names), it's important to clarify the user's intent or default to the more informative option. This ambiguity can lead to outputs that are technically correct but practically unhelpful, affecting the user's ability to interpret the results.

\paragraph{7. Insufficient Handling of Tied Results in Aggregations}

\noindent Question: "Which year has the most number of concerts?"
\begin{lstlisting}[frame=single, breaklines, keepspaces, basicstyle=\footnotesize]
SELECT YEAR FROM concert GROUP BY YEAR
ORDER BY COUNT(*) DESC LIMIT 1;
\end{lstlisting}

\noindent
\begin{minipage}{0.22\textwidth}
\noindent \textbf{Ground-Truth:}
\begin{lstlisting}[frame=single, breaklines, keepspaces, basicstyle=\footnotesize]
Year
2015
\end{lstlisting}
\end{minipage}
\hfill
\begin{minipage}{0.22\textwidth}
\noindent \textbf{Our Result:}
\begin{lstlisting}[frame=single, breaklines, keepspaces, basicstyle=\footnotesize]
Year   count(*)
2014   3
2015   3
\end{lstlisting}
\end{minipage}

\noindent \textbf{Discussion. }
The ground truth SQL does not account for the possibility of ties when determining the year with the most concerts. By applying LIMIT 1, it arbitrarily selects one of the years with the highest count, potentially omitting other equally valid results. Properly handling ties in aggregate functions is essential to provide a complete and accurate answer to the user's query. Adjusting the query to include all years with the maximum number of concerts ensures that the output fully addresses the user's question.

\section{Conclusion}

We proposed the Tree-Driven Sequential Operation QA System (TSO), which transforms natural language queries into logical query plans on structured data without relying on SQL generation. By leveraging large language models (LLMs) to iteratively construct sequences of operations, TSO effectively handles queries of varying complexity. Our experiments on the Spider dataset and a large agronomic dataset with over 8,000 columns demonstrate TSO's ability to process extensive real-world tabular data that many existing QA systems cannot handle. Particularly, TSO successfully manages scientific data with thousands of columns, showcasing its scalability and flexibility. Our work offers a flexible and scalable solution for natural language querying and analysis of large real-world tabular datasets.



\bibliographystyle{ACM-Reference-Format}
\bibliography{citation}

\clearpage
\appendix

\section{Appendix}

\subsection{The NP Hardness or Our Problem}~\label{sec:app_nphard}

We can reduce the classical planning problem to our problem by mapping:

\begin{proof}
We prove that our problem is NP-hard by reducing the Classical Planning Problem to our problem. In the Classical Planning Problem, given an initial state $s_0$, a set of actions $\mathcal{A}$, and a goal state $s_g$, the question is whether there exists a sequence of actions $\pi = \langle a_1, a_2, \dots, a_k \rangle$, where $a_i \in \mathcal{A}$, such that applying $\pi$ to $s_0$ results in $s_g$. We construct an instance of our problem as follows. The dataset $\db$ represents the initial state $s_0$. Each action $a \in \mathcal{A}$ corresponds to an operation $\op_a \in \opset$ that transforms the dataset. The user query $\question$ specifies the goal state $s_g$. If we can find a sequence of operations $p = \langle \op_{a_1}, \op_{a_2}, \dots, \op_{a_k} \rangle$ that, when applied to $\db$, results in a dataset corresponding to $s_g$, then this sequence corresponds to a solution to the Classical Planning Problem. Therefore, solving our problem would solve the Classical Planning Problem. Since the Classical Planning Problem is NP-complete, and we can reduce any instance of it to our problem in polynomial time, the decision problem of our problem is NP-complete. Hence, the optimization version of our problem is NP-hard.
\end{proof}

\subsection{The Agronomic Dataset}~\label{sec:app_GRDC}

The agronomic dataset~\cite{newman2021multiple} is a comprehensive resource designed to monitor and analyze crop growth and development across a range of environmental and meteorological conditions. The dataset is structured around various data domains (DOMs), each offering unique insights into different aspects of the crop trials. These include meteorological data from the Bureau of Meteorology (BOM), satellite-based spectral data, metadata on trials and field management, phenological observations, and environmental variables. Each domain contains columns with structured names that provide a hierarchical description of the variables. The dataset captures time-series data, where many features are recorded at multiple time intervals relative to the planting date. In total, the dataset contains 266033 records and 8058 features.

Domains Overview:
MANDom (Metadata Domain):
The MANDom domain provides metadata related to the crop trials, including information about the trial series, operators, farm machinery, and breeders.

Trial Series: Columns like \nolinkurl{MANDom\_Series\_name} followed by the series identifier (e.g., Durum, Early.Conventional, ITAdvMainLEP) document the trial series names and types.
Breeder Information: Columns such as MANDom\_Breeder followed by the breeder's name (e.g., Advanta.Seeds, Bayer.CropScience) track the entities responsible for breeding the varieties tested in the trials.
Orientation: Columns like MANDom\_Orientation\_sub\_cropNorth, MANDom\_Orientation\_sub\_cropWest indicate the crop orientation for each trial, providing insight into field setup.
PHENDom (Phenological Domain):
The PHENDom domain captures phenotypic observations during crop growth. These include measurements of plant development stages, yield, and other crop characteristics.

Yield Measurements: Columns like \nolinkurl{PHENDom\_yield\_pct\_of\_average}, \nolinkurl{PHENDom\_yield\_t\_ha}track the crop yield either as a percentage of the average or in tons per hectare.
Development Stages: The PHENDom\_Zadoks\_score columns (e.g., PHENDom\_Zadoks\_score\_obs\_1, PHENDom\_Zadoks\_score\_obs\_2) record the Zadoks score at various observations, representing the phenological stages of plant growth.
Grain Quality: Columns like PHENDom\_X1000.grain.weight\_2, PHENDom\_X1000.grain.weight\_6 track important parameters such as grain weight across different measurements.
Disease and Stress Resistance: Columns like \nolinkurl{PHENDom\_Yellow\_Leaf\_Spot}, \nolinkurl{PHENDom\_Waterlogging}, \nolinkurl{and PHENDom\_Weed\_score} indicate the plant's resilience against environmental stressors and disease.
METADom (Metadata Domain for Field and Chemical Management):
The METADom domain provides information on crop and chemical rotations, soil tests, and fertilizer applications.

Crop Rotations: Columns like \nolinkurl{METADom\_Crop\_rotation\_minus\_5\_sub\_cropWheat} and \nolinkurl{METADom\_Crop\_rotation\_minus\_4\_sub\_cropField}. Pea records the sequence of crops grown in previous years, offering insight into crop management practices.
Soil Tests: Columns such as \nolinkurl{METADom\_Soil\_test\_class\_10cm\_Colwell, METADom\_Soil\_test\_class\_60cm\_Bray} report the results of soil tests, providing data on soil properties at different depths.
Previous Crop: Fields like \nolinkurl{METADom\_previous\_crop\_same\_sub\_cropTRUE} indicate whether the same crop was planted in consecutive years, potentially influencing soil health and yield outcomes.
ENVDom (Environmental Domain):
The ENVDom domain contains environmental variables that could impact crop yield, such as damage from pests, animals, or herbicides.

Damage Assessment: Fields like ENVDom\_Damage provide insight into environmental damage affecting crops during the growing season.
BOMDom (Bureau of Meteorology Domain):
The BOMDom domain captures key meteorological data such as temperature and rainfall, tracked over time for each trial.

Temperature: Columns like \nolinkurl{BOMDom\_max\_temperature\_mean\_.80 through BOMDom\_max\_temperature\_mean\_250} and \nolinkurl{BOMDom\_min\_temperature\_mean\_0} through \nolinkurl{BOMDom\_min\_temperature\_mean\_250} provide time-series data of maximum and minimum temperatures for specific days relative to planting (e.g., -80 days before planting to 250 days after).
Rainfall: Similar to temperature, columns like BOMDom\_rainfall\_mean\_0 through BOMDom\_rainfall\_mean\_250 track daily rainfall data.
Other Meteorological Variables: Columns such as BOMDom\_solar\_exposure\_mean provide information on sunlight exposure during the crop's growing season.
SatDom (Satellite Domain):
The SatDom domain includes remote sensing data obtained from satellite observations, capturing a variety of spectral bands and other atmospheric and vegetative properties.

Spectral Data: Columns like \nolinkurl{SatDom\_blue\_band\_mean\_80}, \nolinkurl{SatDom\_NIR\_mean\_70}, \nolinkurl{SatDom\_MIR\_mean\_60} represent reflectance data in different spectral bands (e.g., blue, NIR, MIR), which are important for analyzing vegetation health.
Vegetation Indices: Columns such as \nolinkurl{SatDom\_NDVI\_mean\_80}, \nolinkurl{SatDom\_EVI\_mean\_60}, \nolinkurl{SatDom\_FPAR\_mean\_70} provide indices that track vegetation greenness, photosynthetic activity, and leaf area.
Temperature and Evapotranspiration: Fields like SatDom\_LST\_day\_mean\_80, SatDom\_LST\_night\_mean\_70 track land surface temperature during the day and night, while columns like SatDom\_EvapoTrans\_mean\_80 monitor water loss from crops and soil.
Summary of Data Structure:
Each of the aforementioned domains follows a consistent column-naming convention, which includes a prefix that identifies the data source or domain (e.g., MANDom, PHENDom, METADom, BOMDom, SatDom), followed by a descriptor that provides information on the specific variable being measured, and ending with a time point suffix (for time-series data) that indicates the day relative to planting. This time suffix allows users to track how each variable changes over the crop's growing period. Additionally, certain domains (such as MANDom and METADom) contain metadata that does not vary over time but provides contextual information essential for interpreting the results of the trials.

This dataset is designed to facilitate the analysis of complex environmental and phenotypic factors affecting crop development and can be used to model relationships between environmental conditions and crop performance over time. The combination of meteorological, satellite, and phenotypic data makes the agronomic dataset a rich resource for agricultural researchers aiming to understand and optimize crop yield under varying environmental conditions.

\subsection{The Full Questions for Agronomic Dataset}~\label{sec:app_GRCD_question}

Table~\ref{table:GRDC} summarizes these questions along with their corresponding difficulty levels and results. The outcomes are labelled as T for correct answers, F for incorrect answers, and S for scenarios where the solution could not find an answer within the maximum iteration limit.

\begin{table}[]
\caption{The Experimental Result on the Agronomic Dataset}
\label{table:GRDC}
\small
\begin{tabular}{p{7cm}l}
\hline
Question - Easy Hardness                                                                                                             & Result \\ \hline
1. What is the mean grain yield in tons per hectare?                                                                                  & T      \\
2. What is the maximum recorded maximum temperature on the planting day?                                                              & T      \\
3. List all the trial series names under Early Conventional management.                                                               & T      \\
4. How many trials have a recorded waterlogging score?                                                                                & T      \\
5. What is the average rainfall on the day of planting?                                                                               & T      \\
6. What is the mean NDVI value 10   days after planting?                                                                                & T      \\
7. List the different crop rotations recorded one year before planting.                                                               & F      \\
8. What is the average 1000-grain   weight recorded in the second observation?                                                          & T      \\
9. What is the minimum night-time land surface temperature 20 days after planting?                                                    & T      \\
10. List all the breeders involved in the trials.                                                                                      & T      \\ \hline
Question - Medium Hardness                                                                                                           &        \\ \cline{1-1}
11. For trials where the previous crop was wheat, what is the average grain yield?                                                     & T      \\
12. Calculate the average grain yield for each breeder listed in the dataset.                                                          & T      \\
13. What is the average grain weight for trials with a high waterlogging score?                                                        & T      \\
14. Compare the average EVI values between trials with northern and southern crop orientations.                                        & T      \\
15. How does cumulative evapotranspiration over the first 80 days after planting relate to grain yield?                              & F      \\ \hline
Questions - Hard Hardness                                                                                                            &        \\ \cline{1-1}
16. Perform PCA on the spectral data from satellite observations and identify the top 3 principal components.                          & T      \\
17. Reduce the dimensionality of   METADom using PCA to less than 20 dimensions and provide data for trials with high red band values. & T      \\
18. Predict grain yield using satellite-derived vegetation indices and evaluate the model's accuracy.                                  & T      \\
19. For trials with high waterlogging, assess whether soil properties contribute to the condition.                                     & S      \\
20. Use machine learning to predict the breeder based on phenotypic and environmental data.                                            & S      \\ \hline
\end{tabular}
\end{table}

\subsection{LLM Models and Parameters.}

We utilize several large language models (LLMs) in our experiments to evaluate the performance of our solution:

\textbf{Llama 3.1}: We use the Llama 3.1 model with 70 billion parameters, denoted as \texttt{Llama3.1:70-ins-q4}. This model is known for its strong performance on various language understanding tasks and provides a solid baseline for comparison.

\textbf{GPT-4o-mini}: This is a smaller version of the GPT-4o model, named \texttt{gpt-4o-mini-2024-07-18}. It offers a balance between computational efficiency and performance, making it suitable for testing the scalability of our solution.

\textbf{GPT-4o}: We employ the full GPT-4o model, version \texttt{gpt-4o-2024\\-05-13}, which is a state-of-the-art language model with advanced reasoning capabilities. Its superior performance on complex tasks allows us to assess the upper bounds of our solution's effectiveness.

In our Table-Column Retriever, we set the thresholds $\theta_t = 0.75$, $\theta_c = 0.75$, and $\theta_l = 0.75$, which are used to filter relevant tables, clusters, and columns based on similarity scores between the query embedding and the data embeddings.

We use the embedding model \texttt{thenlper/gte-large} to generate vector embeddings for text descriptions. This model facilitates the retrieval of relevant data elements in our three-level vector index by capturing the semantic meaning of the text.

\end{document}